\documentclass{elsart}
\usepackage{amssymb}
\usepackage{epsfig,graphicx}
\usepackage{graphicx}
\usepackage{amsmath}
\usepackage{epsfig,graphics,color,graphicx,amsmath}
\usepackage{xcolor}
\colorlet{darkred}{red!80!black}
\colorlet{darkgreen}{green!50!black}
\usepackage[normalem]{ulem}

\usepackage{soul}

\usepackage{mathtools}
\usepackage{mathrsfs}
\usepackage{bm}
\usepackage{relsize}
\usepackage{indentfirst}

\colorlet{darkgreen}{green!50!black}
\colorlet{brightyellow}{yellow!75!red}
\colorlet{orange}{red!50!yellow}
\colorlet{darkblue}{blue!60!black}
\colorlet{darkred}{red!80!black}

\newcommand{\cd}{\makebox[0.08cm]{$\cdot$}}

\begin{document}
\begin{frontmatter}
\title{Bound state structure and electromagnetic form factor beyond the ladder approximation}
\author[UNICSUL]{V. Gigante},
\author[ITA]{J. H. Alvarenga Nogueira},
\author[ITA]{E. Ydrefors},
\author[IFT]{C. Gutierrez}, 
\author[LBD]{V.A. Karmanov}
and
\author[ITA]{T. Frederico}
\footnote{Corresponding author: tobias@ita.br}
\address[UNICSUL]{Laborat\'orio de F\'\i sica Te\'orica e 
Computacional - LFTC, Universidade Cruzeiro do Sul, 01506-000 S\~ao Paulo, Brazil} 
\address[ITA]{Instituto Tecnol\'ogico de Aeron\'autica, DCTA,
12228-900,\\ S\~ao Jos\'e dos Campos,~Brazil}
\address[IFT]{Instituto de F\'\i sica Te\'orica, UNESP,
01156-970, S\~ao Paulo, Brazil}
\address[LBD]{Lebedev Physical Institute, Leninsky Prospekt 53, 119991 Moskow, Russia}

\date{\today}

\maketitle

\begin{abstract}
We investigate the response of the bound state structure of a two-boson system,
within a Yukawa model with a scalar boson exchange, to the inclusion of 
the cross-ladder contribution to the ladder kernel of the Bethe-Salpeter equation.
The equation is solved by means of the Nakanishi integral representation and light-front 
projection.
The valence  light-front wave function and the elastic electromagnetic form factor 
beyond the  impulse approximation, with the inclusion of the two-body current, generated by the cross-ladder kernel, are computed. 
The valence wave function and electromagnetic form factor, 
considering both ladder and ladder plus cross-ladder kernels, are studied in detail. Their asymptotic forms are found to be
quite independent of the inclusion of the cross-ladder kernel, for a given binding energy. The asymptotic decrease of form factor agrees with the counting rules. 
This analysis can be generalized to fermionic systems, with a wide application in the study of the meson structure.
\end{abstract}
\end{frontmatter}

The investigation of fundamental interactions faces the challenge to obtain theoretically the 
properties of relativistic bound systems in the Minkowski space. One relevant present example, 
 is the introduction of quasi-parton distributions calculated with moving hadrons 
in the Euclidean Lattice QCD for large longitudinal momentum to match with parton distribution functions (PDFs) 
in the infinite momentum frame \cite{XJiPRL13}. 
On the other side, recent tools are being introduced to investigate the spectrum and the Minkowski space structure of 
composite systems  within the continuum approach to bound states in field theory, without resorting to the Wick rotation
in the Bethe-Salpeter (BS) equation. One technique to solve  bound and scattering problems within the BS approach relies
on the Nakanishi integral representation (NIR) \cite{nakanishi}. This method was introduced about two decades ago in
\cite{KusPRD95}, and further developed in \cite{KarEPJA06,CarEPJA06} where the projection onto the light-front (LF) was 
used as an  essential step to simplify the formalism. Later on, it was further extended to scattering states \cite{FrePR12}, and 
computations using convenient polynomial basis expansion were provided in \cite{FrePRD14,FreEPJC15}. 
The efforts were undertaken  to invert the NIR for the Euclidean BS amplitude, in order to find the Nakanishi weight function  
and use it to reconstruct the BS amplitude in Minkowski space~\cite{FCGK_LC_2015}.
Applications to bound fermionic systems were also done \cite{CarEPJA10,dPaPRD16}. The method has been shown to be 
reliable to study the spectrum and the Minkowski space structure of a relativistic two-boson system in the ladder 
approximation \cite{GutPLB16}. Impact parameter space amplitudes (see e.g. \cite{BurIJMPA03}) can be derived from the 
LF valence wave function of the ground and excited states.
Together with the asymptotic form of the LF wave function for large transverse momentum, 
the impact parameter 
space representation at large distances were studied within the ladder approximation in \cite{GutPLB16}.
The Minkowski space approach has been extended beyond the ladder exchange in Ref. \cite{CarEPJA06}, where it 
was considered the cross-ladder contribution to the kernel of the BS equation for the two-boson bound state. 
The binding energy as a function of the coupling constant
was also computed in the Euclidean space approach within the Feynman-Schwinger framework of the Yukawa model for 
two-boson bound states \cite{NiePRL96}, where all possible cross-ladder diagrams were taken into account. This work shows quite clearly the extra attraction provided by adding the infinite set of diagrams in the kernel of the corresponding BS equation. 
Indeed, the consideration of only lowest order cross-ladder diagram gives a considerable
net attraction in the two-boson bound state (see e.g. \cite{CarEPJA06}).  Therefore, it is natural to expect that 
the dynamics beyond the ladder exchange is reflected not only on the binding energy but also on the Minkowski space structure of the
bound state. This
is modified by the higher order contributions to the kernel of the BS equation, even if the binding energy is kept 
fixed by changing the coupling constant. 

The interesting question comes on how the dynamics beyond the ladder exchange contributes quantitatively to the
asymptotic behavior of both valence LF wave function, associated with the PDFs, and to
the elastic electromagnetic (EM) structure. In this paper we study the two-boson bound state structure in the Yukawa model,
considering the ladder and ladder plus cross-ladder kernels, to investigate quantitatively the LF wave function and
elastic EM form factor. To satisfy the gauge invariance, the cross-ladder graph must also contribute to the EM current 
of the bound pair as a two-body current, 
which is also considered in this work. These observables are intrinsically connected with the Minkowski space structure of the
 bound state. 
The solution of the homogeneous two-boson BS equation in Minkowski space is found numerically by transforming 
it into a non-singular integral equation 
for the weight function provided by the NIR of the BS amplitude, using the technique proposed in 
\cite{KarEPJA06,CarEPJA06} and further developed in \cite{FrePRD14}. 

To put our work on a broad perspective, we recall that in the realm of the 
quark counting rules \cite{MatLNC73,BrodPRL73} within
perturbative QCD, it was derived  the leading asymptotic large momentum form of amplitudes for 
exclusive processes, in particular to elastic form factors \cite{LepPRD80,BrodPRL85}. Higher-twist contributions 
to the associated amplitudes are subleading \cite{LepPRD80}.
These ideas applied to a spin 1 two-fermion bound state resulted in the "universal ratios" \cite{Brodsky:1992px} 
between the leading asymptotic contributions  to the elastic EM form factors. 
Later on, subleading power corrections were considered in the EM form factors  of the deuteron  \cite{KobPRD94,KobPAN95} and   
$\rho$-meson \cite{MelPLB16}, and in exclusive processes \cite{Carson2003} consistent with the LF angular conditions.
Furthermore, the quark counting rules were generalized in Ref. \cite{XJiPRL03} for the leading hard transverse momentum 
dependence of the Fock components of the hadronic LF wave function in terms of the  parton number, 
orbital angular momentum along the z-direction and hadron helicity. 
The present study is a preparation for future applications to explore the nonperturbative physics of QCD. We address
the issue on how the asymptotic behavior from counting rules is formed qualitatively and quantitatively, considering the
NIR of the BS amplitude,
both in the valence LF wave function and EM form factor. In addition, we study quantitatively 
the elastic two-body current associated with a cross-ladder term being a higher-twist contribution 
to the form factor. Our aim is to determine 
how it is damped with respect to the leading term in the present nonperturbative calculation of the bound state. Another aspect
analyzed here is the question on how the asymptotic behavior of the form factor and LF wave function change
with the  modified kernel. For this study we consider a 
fixed binding energy, which keeps the low momentum behavior of these quantities quite independent of the kernel choice, allowing us to
focus on the high momentum region independent on the binding energy.

This work is organized as follows. In Sect. \ref{sect:BSNIR}, the BS equation and NIR 
of the BS
amplitude are briefly introduced. In Sect. \ref{sect:LFWF} the valence LF wave function is studied in detail 
with respect to its asymptotic form and the role of the ladder exchange in forming the leading large momentum behavior.
In Sect. \ref{sect:SLEMFF} the space-like EM form factor is introduced and the current conservation is discussed. 
The numerical results for the impulse  and two-body current contributions to the form factor are presented, and a discussion
of the ladder exchange dominance is performed quantitatively. The asymptotic  behavior of the form factor is derived
in Sect. \ref{sect:ASYFF}, and we illustrate them numerically. In Sect. \ref{sect:SUMOUT}, we provide the
summary  and an outlook for future developments of our work.

\section{ Bethe-Salpeter Equation and Nakanishi Integral Representation} \label{sect:BSNIR}

The BS equation in Minkowski space, for two spinless particles,
reads:
\begin{equation}\label{bs}
\Phi(k,p)=S\left(\frac{p}{2}+k\right)\,S\left(\frac{p}{2}-k\right)
\int \frac{d^4k'}{(2\pi)^4} \, iK(k,k',p)\Phi(k',p) \, ,
\end{equation}
where the Feynman propagator is $S(k)=i \left[k^2-m^2+i\epsilon\right]^{-1}$.
The interaction kernel $K$ is given by the sum of irreducible Feynman
diagrams. The ladder kernel  is considered in most of the works,
but here we incorporate also the cross-ladder contribution. 

The BS amplitude is found in the form of the NIR \cite{nakanishi,KusPRD95}:
\begin{equation}
\label{bsint}
\Phi(k,p)=-{i}\int_{-1}^1dz\int_0^{\infty}d\gamma 
\frac{g(\gamma,z)}{D^3(\gamma,z;k,p)}\, ,
\end{equation}
where the Nakanishi denominator is:
\begin{equation}\label{dennaka}
D(\gamma,z;k,p)=\gamma+m^2-\frac{1}{4}M^2-k^2-p\cdot k\; z-i\epsilon \, .
\end{equation}
The weight function $g(\gamma,z)$ itself is not singular, whereas
the singularities of the BS amplitude are fully reproduced by this
integral. The BS amplitude in the form (\ref{bsint}) is substituted into the BS
equation (\ref{bs}) and after some mathematical transformations~\cite{KarEPJA06}, one obtains the following integral equation for
$g(\gamma,z)$:
\begin{small}
\begin{equation} \label{bsnew}
\int_0^{\infty}\frac{g(\gamma',z)d\gamma'}{\Bigl[\gamma'+\gamma
+z^2 m^2+(1-z^2)\kappa^2\Bigr]^2} =
\int_0^{\infty}d\gamma'\int_{-1}^{1}dz'\;V(\gamma,z,\gamma',z')
g(\gamma',z'),
\end{equation}
\end{small}
where for bound states $\kappa^2 = m^2- \frac{1}{4}M^2 > 0$ and $V$ is expressed via the kernel $K$.
The ladder and ladder plus cross-ladder kernels in Eq.~(\ref{bsnew}) were worked out in detail in Refs. \cite{KarEPJA06} and 
\cite{CarEPJA06}, respectively.  The numerical method to solve Eq.~(\ref{bsnew}) was described in detail in \cite{FrePRD14,GutPLB16}, where
it was proposed a basis expansion in Laguerre polynomials for the noncompact variable and Gegenbauer polynomials 
for the compact one.

Noteworthy to point out that the $s-$wave valence LF wave function is written in the form (we follow the convention of our previous paper \cite{FCGK_LC_2015}):
\begin{equation} \label{LFWF}
\psi_{LF}(\gamma,\xi)=\frac{1-z^2}{4 }\,\int_0^{\infty}\frac{g(\gamma',z)d\gamma'}{\Bigl[\gamma'+\gamma
+z^2 m^2+(1-z^2)\kappa^2\Bigr]^2} \, ,
\end{equation}
where the transverse momentum is $k_\perp=\sqrt{\gamma}$ and the LF momentum fraction is $\xi=(1-z)/2$ with $0<\xi<1$. 
The physical or normal solutions of the BS equation are the ones for which the weight function has the symmetry property 
$g(\gamma,z)=g(\gamma,-z)$, and it is reflected in the expected symmetry of the valence wave function for two identical bosons.

\begin{table}[h!]
  \centering
  \begin{tabular}{cccccc}
\hline\hline  
    $B/m$ & $\mu/m$ & $\alpha^{(L+CL)}$& $\alpha^{(L)}$& $\alpha^{(L)}/\alpha^{(L+CL)}$& $\psi^{(L)}_{LF}/\psi^{(L+CL)}_{LF}$ \\
\hline\hline  
1.5 & 0.15 & 4.1399 & 6.2812 & 1.5172 & 1.5774 \\
    & 0.50 & 5.1568 & 7.7294 & 1.4988 & 1.5395 \\
\hline
1.0   & 0.15 & 3.5515 & 5.3136 & 1.4961 & 1.5508 \\
    & 0.50 & 4.5453 & 6.7116 & 1.4766 & 1.5094 \\
\hline
0.5 & 0.15 & 2.5010 & 3.6106 & 1.4436 & 1.4805 \\
    & 0.50 & 3.4436 & 4.9007 & 1.4231 & 1.4405 \\
\hline
0.1 & 0.15 & 1.1052 & 1.4365 & 1.2997 & 1.2763 \\
    & 0.50 & 1.9280 & 2.4980 & 1.2956 & 1.2694 \\
\hline\hline  
  \end{tabular}
  \caption{ Comparison between the ratio of the coupling constants, given in terms of  $\alpha=g^ 2/(16\pi m^2)$, corresponding
to ladder (L) and ladder plus cross-ladder (L+CL) kernels, with the ratio of the LF
wave functions in the asymptotic limit, namely for a large value of $\gamma=\,500m^2$, and with the particular choice $\xi$ = 1/2. 
For this analysis we use the normalization $\psi^{(L)}_{LF}(0,1/2)=\psi^{(L+CL)}_{LF}(0,1/2)=1$. 
}\label{tab:table1}
\end{table} 

\section{Valence light-front wave function }\label{sect:LFWF}

The cross-ladder kernel is 
attractive as it is known \cite{CarEPJA06}, and therefore  the coupling constant decreases to keep the 
same binding energy, as illustrated by the values of $\alpha^{(L)}$ and $\alpha^{(L+CL)}$ presented in Table~\ref{tab:table1}
for a given $B$ and $\mu$. 
The momentum dependence of the valence wave function is discussed in what follows. We choose 
a strongly bound situation with $B=1.5\,m$ to show the effect of  changing  the interaction kernel from 
ladder to ladder plus cross-ladder at a fixed binding energy.

\begin{figure}[!hbt]
\centering
\includegraphics[scale=0.32,angle=0]{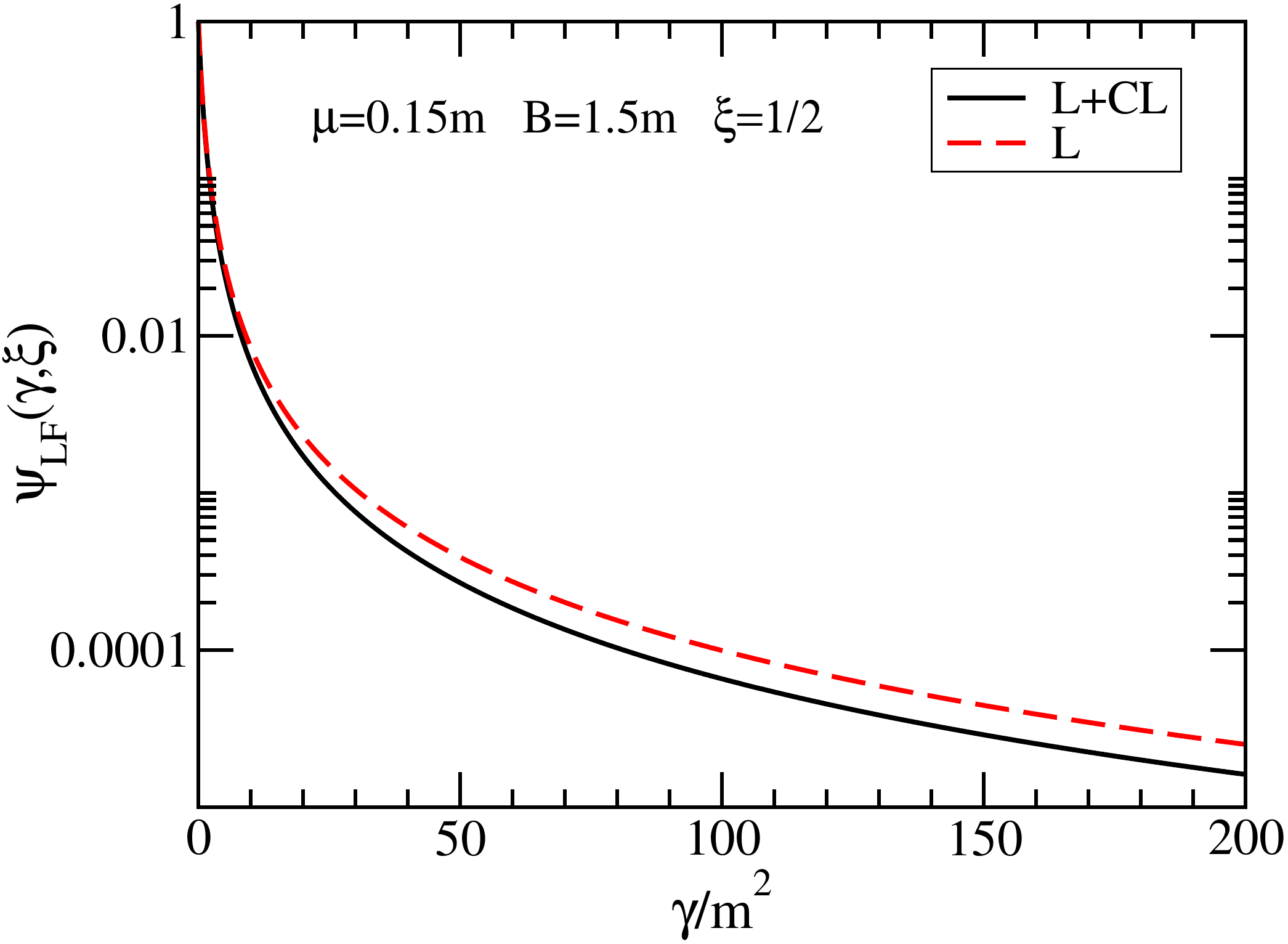}
\includegraphics[scale=0.32,angle=0]{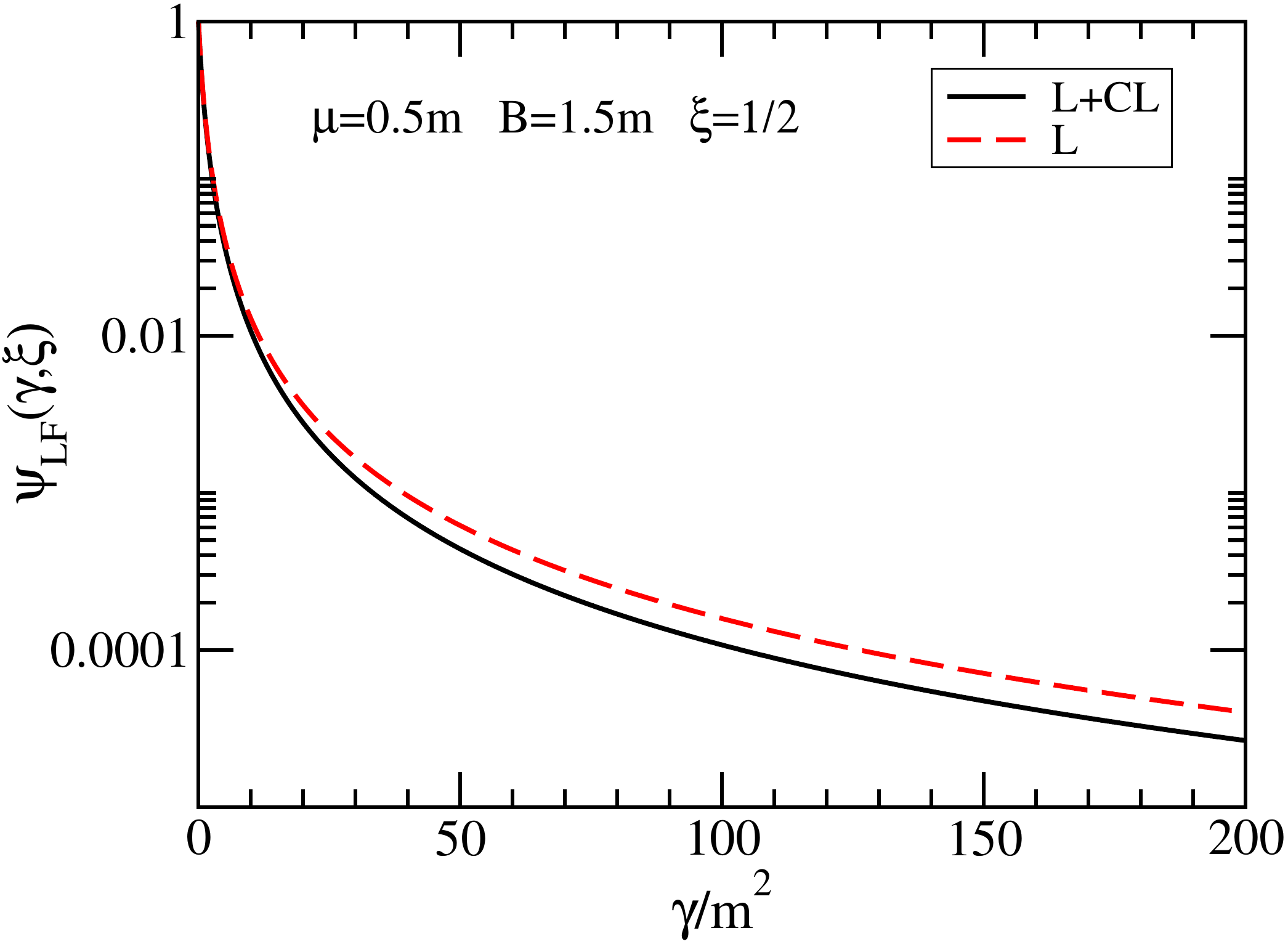}
\caption{LF wave function vs. $\gamma$ for $\xi=1/2$ with ladder (L) (dashed lines) and ladder plus cross-ladder (L+CL)
(solid lines) interaction kernels for  $B=1.5\, m$ and $\mu=0.15\,m$ (left-frame) and ${\mu=} 0.5\,m$ (right-frame). } \label{fig:wflf1}
\end{figure}

\begin{figure}[!hbt]
\centering
\includegraphics[scale=0.32,angle=0]{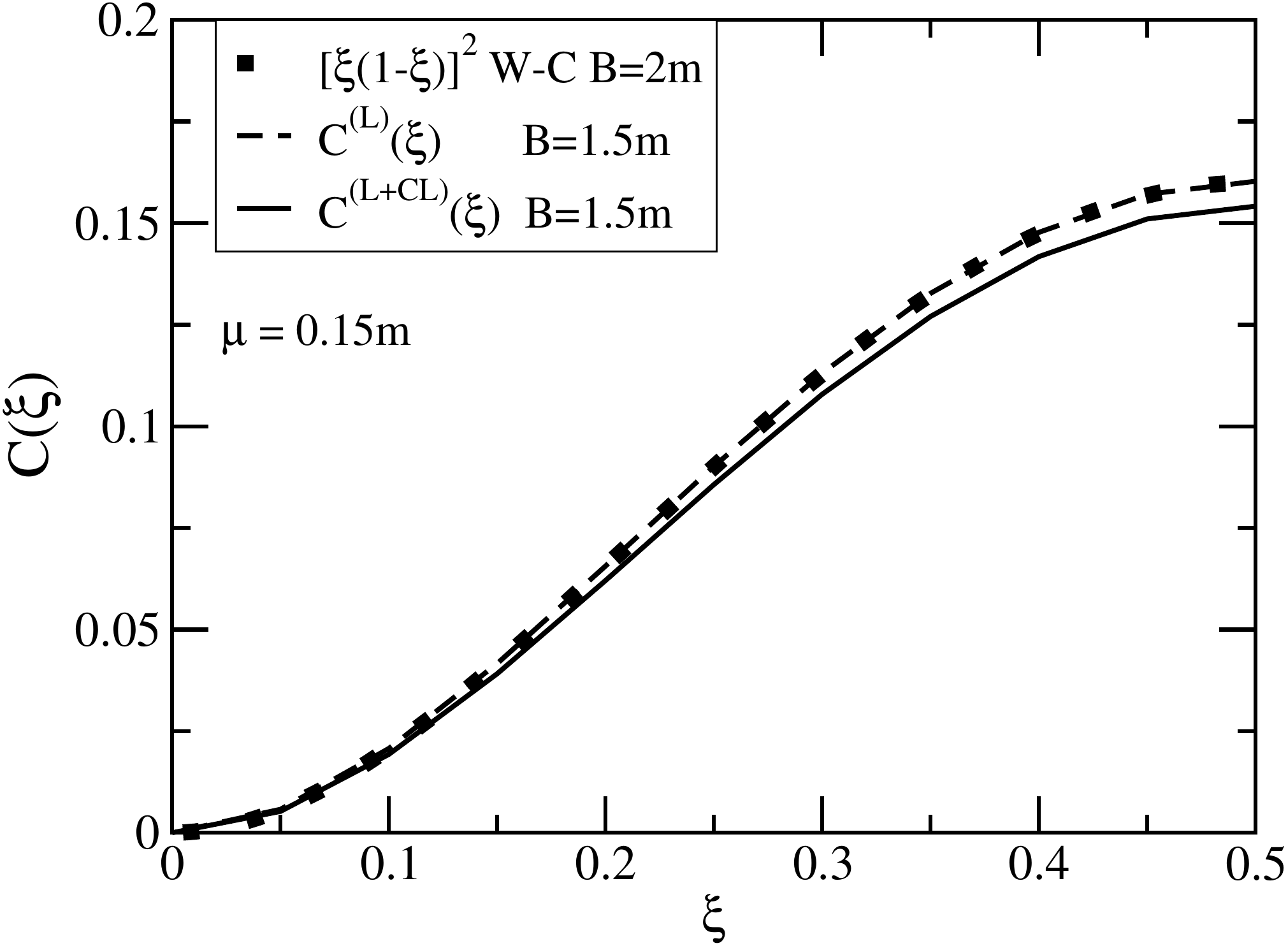}\includegraphics[scale=0.32,angle=0]{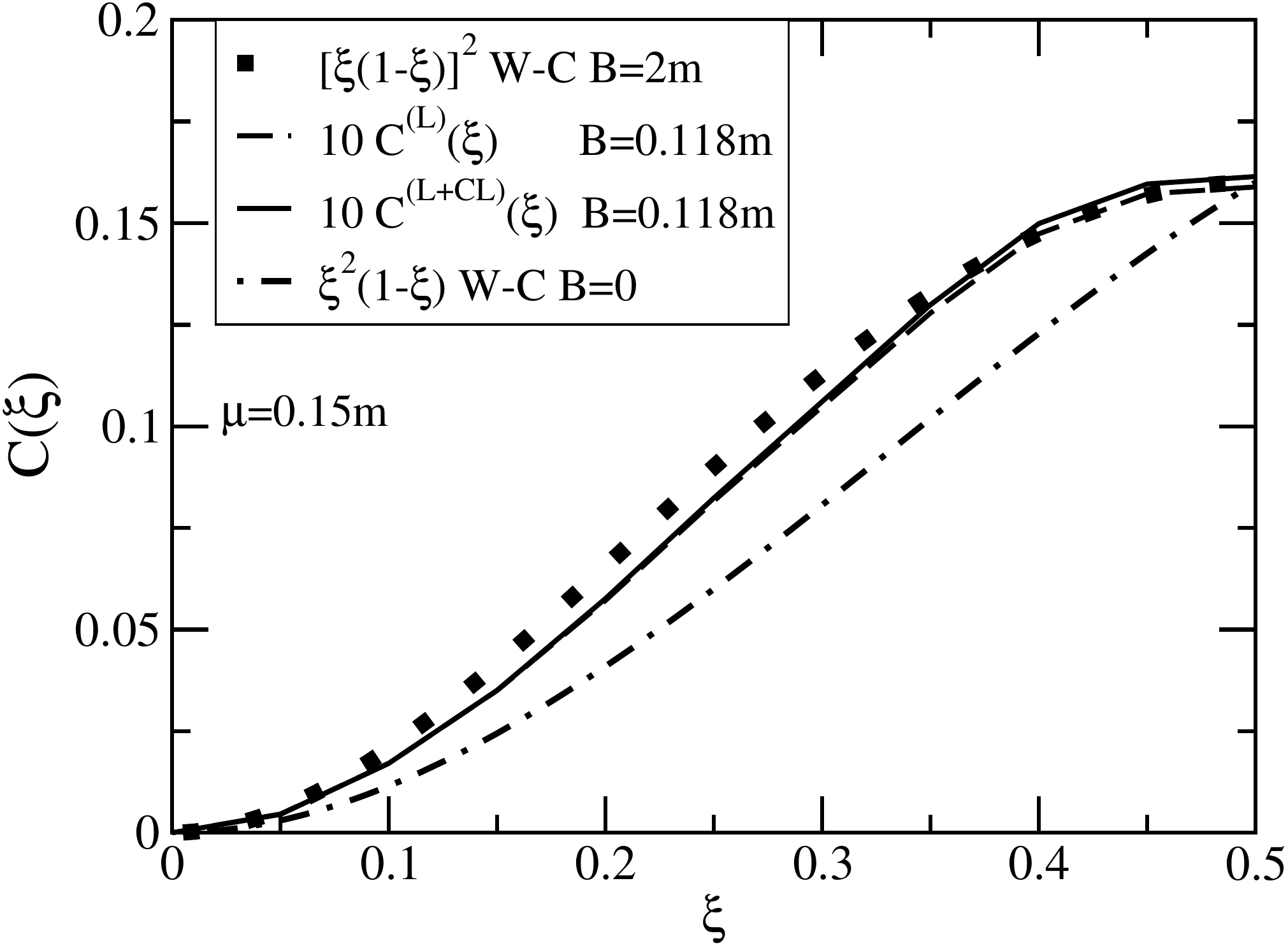}
\caption{ Asymptotic function $C(\xi)$ defined from the LF wave function for $\gamma\to\infty$ (\ref{cxi}) computed for the ladder kernel, 
$C^{(L)}(\xi)$ (dashed line), and ladder plus cross-ladder kernel, $C^{(L+CL)}(\xi)$ (solid line),  with exchanged boson mass of $\mu=0.15\,m$. Calculations are performed for $B=1.5\, m$ (left frame) and  $B=0.118\, m$ (right frame). A comparison with the analytical forms of $C(\xi)$ valid for the Wick-Cutkosky model for $B=2m$ (full box) and $B\to 0$ (dash-dotted line) both arbitrarily normalized.} \label{fig:cxi}
\end{figure}

The result for the wave function is  shown in Fig.~\ref{fig:wflf1}.
At relatively low momentum, $\sqrt{\gamma}\lesssim 3\, m$, the wave function is practically the same 
for the ladder and ladder plus cross-ladder kernels. That happens because this momentum region is determined by 
the binding energy that gives the behavior of the wave function at large distances. 
In the present case one should expect that the momentum region determined mainly by
the binding energy is of the order of ${\sqrt{\gamma}\sim}B= 1.5 \, m$, which seems to be the case. 
At large momentum, we observe that the ladder and ladder plus cross-ladder results for the wave 
function, are essentially proportional. According to the general discussion on asymptotic 
behavior of the LF wave function \cite{LepPRD80}, the large momentum tail should be 
dominated by the ladder exchange, that is common to both calculations. In \cite{GutPLB16},
it was found, for ground and excited states, that for $\gamma\to \infty$:
\begin{equation}\label{cxi}
\psi_{LF}(\gamma,\xi) \to  \alpha \ \gamma^{-2}\,C(\xi) \, ,
\end{equation}
where  $\alpha$ is factorized and $\psi_{LF}(0,1/2)=1$ is chosen to get $C(\xi)$ in Fig.~\ref{fig:cxi}. 

The only fact that the kernel can be enlarged to include 
the cross-ladder, allows us to check how  $C(\xi)$ changes for a given binding energy, considering 
that the coupling constant has to be modified for the two kernels to keep $B$ fixed and the ladder exchange
dominates the large momentum region. Table~\ref{tab:table1} illustrates  how the asymptotic wave function scales with $\alpha$ 
for ladder and ladder plus cross-ladder  kernels with $\mu=0.15 \, m$ and $0.5 \, m$. We considered 
the coupling constants for different binding energies, and
the ratio of the wave functions ($\psi^{(L)}_{LF}/\psi^{(L+CL)}_{LF}$) when $\gamma=500\,m^2$ 
and $\xi=1/2$ ($z=0$). The Table also illustrates that the ratio between the values of $\alpha$
are about the same as the ratio of the wave functions, namely $\alpha^{(L)}/\alpha^{(L+CL)}\approx\psi^{(L)}_{LF}/\psi^{(L+CL)}_{LF}$. 
This turns clear the motivation in factorizing $\alpha$ in Eq.~(\ref{cxi}).

The asymptotic form in Eq.~(\ref{cxi}) is also found in the Wick-Cutkosky (WC)  model, 
where the valence ground state wave function is \cite{HwaNPB04}:
\begin{equation}\label{lfwvwc}
\psi^{(WC)}_{LF}(\gamma,\xi)=\frac{C^{(WC)}(\xi)}{2\sqrt{\pi}(\gamma+m^2-\xi(1-\xi)M^2)^2}
\end{equation}
with $C^{(WC)}(\xi)=\xi(1-\xi)g^{(WC)}(1-2\xi)$. In the two extreme limits of binding energy, strongly and weakly 
bound state, this function is found analytically and it is given by
\begin{equation}\label{wcs}
C^{(WC)}(\xi)=[\xi(1-\xi)]^2\, ,
\end{equation}
for $B=2\:m$, and by,
\begin{equation}\label{wcw}
C^{(WC)}(\xi)=\xi(1-\xi)\left(\frac12-\big|\frac12-\xi\big|\right)
\end{equation}
for $B\to 0$. The normalization of $C^{(WC)}(\xi)$ presented above is chosen arbitrarily.

The asymptotic functions $C(\xi)$ defined from the LF wave function for $\gamma\to\infty$ (\ref{cxi}) obtained with the ladder
and ladder plus cross-ladder kernels are shown in Fig.~\ref{fig:cxi}, for weak and strong binding energies  $B=0.118\,m$ and $B=1.5\,m$,
respectively. We choose the case of an exchanged boson mass of $\mu=0.15\,m$. As mentioned, for this study 
the normalization of the wave function is chosen as $\psi_{LF}(0,1/2)=1$.
First we observe, a quite weak sensitivity in the form of $C(\xi)$ with $B$, for
the values we use, while the Wick-Cutkosky model in the extreme limits of binding energy has $C(\xi)$ quite different as given by 
Eqs.~(\ref{wcs}) and (\ref{wcw}). The noticeable difference in the weak and strong binding cases is the magnitude of $C(\xi)$,
which from $B=1.5\,m$ to $0.118\,m$ decreases by a factor of 10, considering that the normalization  $\psi_{LF}(0,1/2)=1$ 
is fixed in both cases.  
This is the expected behavior as the wave function in the
strong binding case spreads out for larger momentum than in the weak binding situation. Our results are closer
to the analytical form of $C(\xi)$ obtained in the Wick-Cutkosky model for $B=2m$. This comparison suggests that
$C(\xi)$ is well approximated by $[\xi(1-\xi)]^\lambda$ with $\lambda$ close to 2 for small $\mu$. In the extreme case of 
$\mu=\infty$ the asymptotic form of the LF wave function changes  to $\gamma^{-1}$, while $c(\xi)=[\xi(1-\xi)]^2$.
 
 We remind that the end-point behavior of the LF wave function is immediately associated by Eq.~(\ref{LFWF}) with the behavior of the 
 Nakanishi weight function $g(\gamma,z)$ at $z\to\pm 1$. The quadratic form at the end point of $C(\xi)$ comes from a linear 
 damping of $g(\gamma,z)\sim (1-|z|)$ for $|z|\to 1$. This property will be later on used to study analytically 
 the asymptotic form of the EM form factor and show the consistence of the formulation with the counting rules.

Our study of the structure of the bound state continues now with the analysis of the elastic EM form factor.
We check the effect of the addition of the cross-ladder to the kernel, by  comparing results with a fixed binding energy. 
We explore the low and high momentum transfer regions, with the aim to verify the asymptotic behavior and the
dominance of the ladder exchange for large momentum. Furthermore, the ladder plus cross-ladder kernel offers the opportunity
to study the effect of the two-body  current in the form factor and we show analytically and 
quantitatively its faster decay with momentum transfer, as it is expected from a higher-twist contribution to the form factor
\cite{LepPRD80}.

\section{Space-like Electromagnetic Form factor}\label{sect:SLEMFF}

For a spinless system in the general case the e.m. current (not necessary elastic and conserved) is given by
\begin{equation}\label{current}
J_\mu = (p_\mu+p'_\mu) F_{1}(Q^2) + (p_\mu-p'_\mu) F_2(Q^2)\, ,
\end{equation}
where $Q^2=-(p-p')^2>0$. 
In the elastic case current conservation implies that $F_2=0$ and only $F_1$ survives
and represents the virtual photon absorption amplitude by the composite system. 

\begin{figure}[htb!]
\begin{center}
\includegraphics[scale=.8]{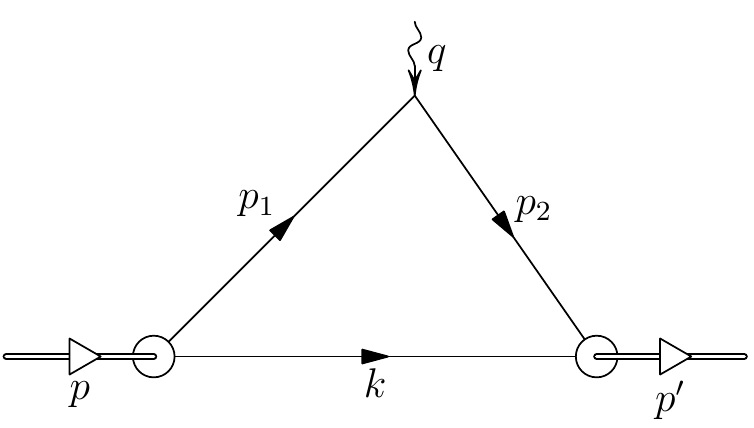}
\includegraphics[scale=0.8]{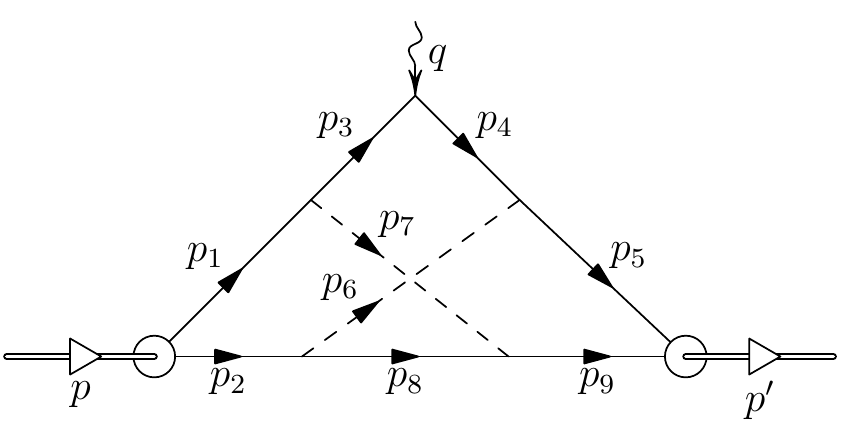}
\caption{Diagrammatic representation of the photon absorption amplitude: impulse (left) and two-body current contribution (right).}\label{triangle}
\end{center}
\end{figure}

For our kernel considered up to the cross-ladder the gauge invariance of the EM coupling implies two
irreducible contributions to the photon absorption amplitude, which leads to two parts of the form factors
\begin{equation}
F_{1}(Q^2)=F_{I}(Q^2)+F_{X}(Q^2),
\end{equation}
where $F_{I}$ means the impulse contribution, obtained from the triangle diagram, Fig.~\ref{triangle} (left),
and $F_{X}$  is the two-body current contribution to the form factor,
which is computed from the virtual photon absorption amplitude diagrammatically depicted in Fig.~\ref{triangle} (right).

\subsection{Impulse contribution to the form factor} 

The impulse contribution to the form factor is represented diagrammatically in Fig.~\ref{triangle}.
For a system composed of two spinless particles, the EM vertex can be expressed in terms of the BS amplitude by the formula
\begin{equation}\label{ff}
(p+p')^{\mu} F_{I}(Q^2) =
i\int \,\frac{d^4k}{(2\pi)^4} (p+p'-2k)^{\mu}\,(k^2-m^2)\,
\Phi \left(\frac{p}{2} -k,p\right)\Phi \left(\frac{p'}{2} -k,p'\right).
\end{equation}
We contract both sides of (\ref{ff}) with $(p+p')_{\mu}$ and
substitute in its r.h.-sides the BS amplitude in terms of the NIR given in Eq.~(\ref{bsint}):
\begin{multline}\label{J}
F_{I}(Q^2)=\frac{i }{(2\pi)^4}
 \int_{0}^\infty d\gamma \int_{-1}^1 dz\,\int_{0}^\infty d\gamma'\int_{-1}^1dz'\,\int d^4k\left[1-\frac{2k\cdot (p+p')}{(p+p')^2}\right]\\ \times
\frac{(m^2-k^2) \,g(\gamma,z)g(\gamma',z')}
{D^3(\gamma,z;\frac{p}{2}-k,p)\,D^3(\gamma',z';\frac{p'}{2}-k,p')}\, ,
\end{multline}
$D$ is defined in (\ref{dennaka}).
The loop integral in $d^4k$ is  calculated analytically by means of the Feynman parametrization. 
This procedure is described in detail in Ref.~\cite{CarEPJA09}. In this way,  one finds the exact formula in terms of the weight function
$g(\gamma,z)$:
\begin{small}
\begin{multline}\label{ffM}
F_{I}(Q^2)=\\=\frac{1}{ 2^7\pi^3}\int_0^\infty d\gamma\int_{-1}^1
dz\, g(\gamma,z) \int_0^\infty d\gamma' \int_{-1}^1 dz'\,
g(\gamma',z')
\int_0^1 dy\,y^2(1-y)^2 \frac{f_{num}}{f_{den}^4},
\end{multline}
\end{small}
where 
\begin{multline}\label{C}
f_{num}=(6 \eta-5)m^2 + [\gamma' (1 - y) + \gamma y] (3 \eta -2) 
+ 2M^2 \eta(1-\eta) +\\ + \frac{1}{4} Q^2 (1 - y) y (1+z) (1+z')
\\
f_{den}= m^2 + \gamma' (1 - y) + \gamma y -
            M^2 (1 - \eta ) \eta
+ \frac{1}{4}Q^2 (1 - y) y (1+z) (1+ z'),
\end{multline}
with 
$
2\,\eta=(1 + z)y + (1+z')(1-y).
$

\subsection{Two-body current contribution to the form factor}

Next, we sketch the computation of the form factor for the two-body current represented by the 
diagram shown in the right of Fig.~\ref{triangle}, where the photon vertex is  given by $-i(p_4+p_3)^{\mu}$.
The form of the two-body current in terms of the BS amplitudes of the final and initial state is written as:
\begin{multline}\label{formfac-2}
 F_{X}(Q^2)= - i \frac{g^4}{ (2 \pi)^{12} }\int d^4p_{2} d^4p_{8} d^4p_{9}
 \, \left[1- 2\frac{ (p+p')\cdot(p_9 + p_2 -p_8)}{(p+p')^2} \right]
\\ 
\times \left[\prod_{i=3,\,i\neq 5}^8\frac{1}{p_i^2-m_i^2+i\epsilon}\right]  \,\Phi \left(\frac{p}{2}-p_2,p\right)\, \Phi \left(\frac{p'}{2}-p_9,p'\right)\, \, ,
\end{multline}
where $p_3=p-p_9-p_2+p_8$, $p_4=p'-p_9-p_2+p_8$, $p_6=p_2-p_8$, $p_7=p_9-p_8$, $m_3=m_4=m$ and $m_6=m_7=\mu$.

After substituting the BS amplitude by the NIR, Eq.~(\ref{bsint}), in the above formula, and using six Feynman parametric integrations, only one 
denominator remains, and by standard integrations over the three loops one obtains
\begin{multline}\label{formfac-5}
F_{X}(Q^2) =  -\frac{3 \alpha^2 m^4}{(2 \pi)^5} \int_0^\infty d\gamma  \int_{-1}^1 dz \int_0^\infty d\gamma'\int_{-1}^1 dz'  g(z',\gamma') g(z,\gamma)\\ \times
 \prod_{i=1}^6\int_{0}^{1} dy_i \Theta\left(1-\sum_{j=i+1;i<4}^4y_j\right)
(1-y_5)^2 y_5^2  (1-y_6)^2 y_6^3 \frac{f^X_{num}}{\left[f^X_{den}\right]^5}, 
\end{multline}
where the functions  $f^X_{num}$ and $f^X_{den}$, depends on the $m$, $y_i$, $\gamma$, $z$, $\gamma'$, $z'$, $p'$ and $p$. They do not
contain any singularity, but are too lengthy to be explicitly shown here. For the calculation of the form factor
the above formula is used.

{\it Current conservation.}
The expression for the elastic EM vertex is symmetric relative to the permutation $p \leftrightarrow p'$ both 
for the impulse as well as for the two-body current contributions.
Hence, the second (antisymmetric) term in (\ref{current}) cannot appear in the elastic EM vertex, and
therefore $F_2(Q^2) \equiv 0$. That follows from the contraction of the EM vertices associated with the impulse and two-body current 
terms, diagrammatically shown in Fig.~\ref{triangle}, with $(p-p')^{\mu}$, which results in
zero for any BS amplitude in the elastic case. In this case current conservation is automatically fulfilled 
for any particular contribution to the current.

However, $J_\mu $ is an operator and the current conservation $J\cdot q=0$ means that all the matrix elements of 
this operator must be zero. What we considered for elastic form factor is only one (diagonal) matrix element. The
non-diagonal (transition) matrix elements bound $\to$ excited state also must be zero. 
The above symmetry will not hold in this case and the zero value of $J\cdot q$ should appear as a subtle cancellation of 
different contributions both in the kernel and in the EM vertex. This  cancellation found numerically would be indeed a 
powerful test, as in Ref. \cite{CKtransit}, where this cancellation was demonstrated numerically for the transition form factor
associated with the EM breakup process: bound $\to$ scattering state. In the present work we restrict ourselves by the elastic case only. 
The inelastic transitions and the current conservation in this case will be a subject of forthcoming paper.

\vspace{1cm}
\begin{figure}[!hbt]
\centering
\includegraphics[scale=0.33]{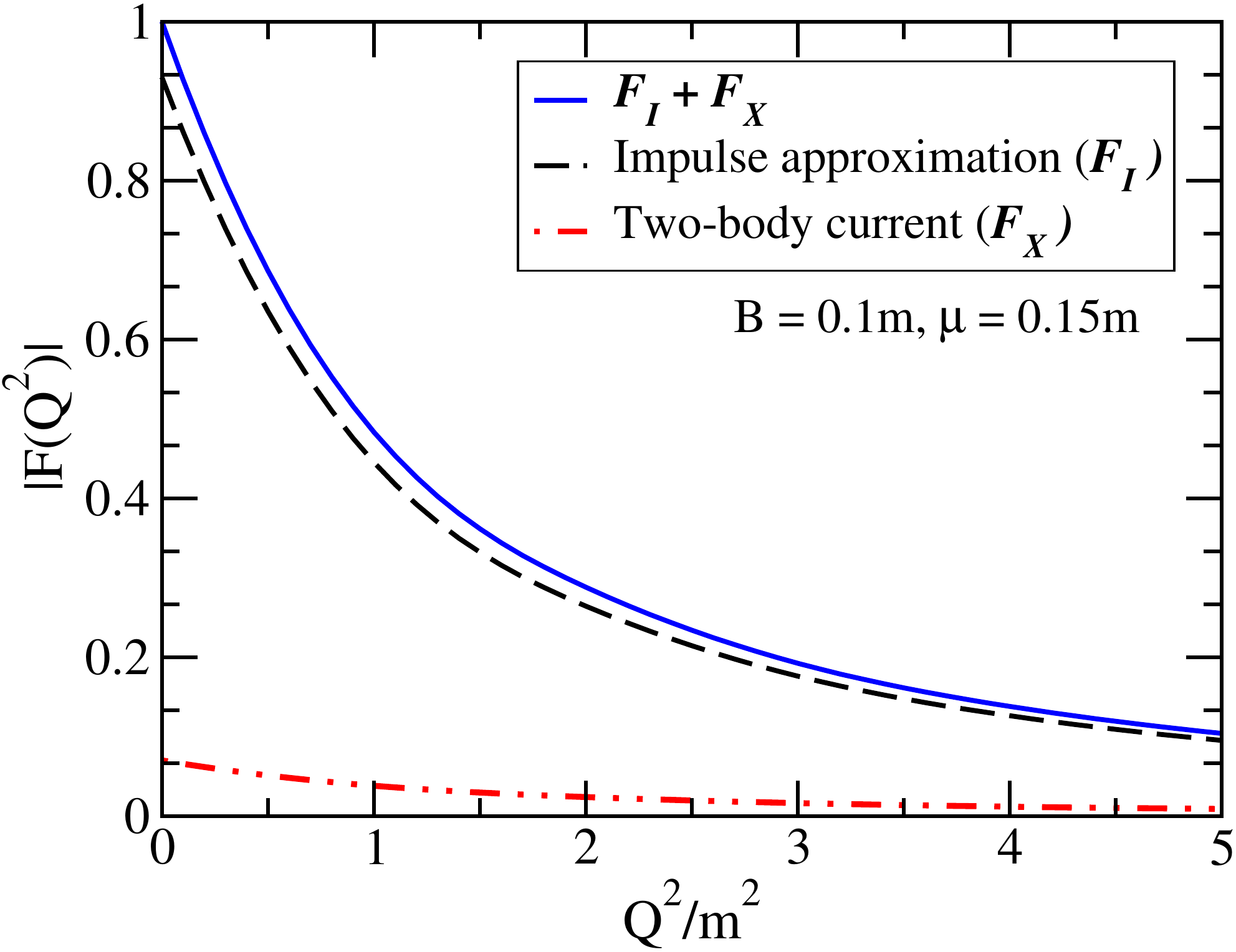}
\includegraphics[scale=0.33]{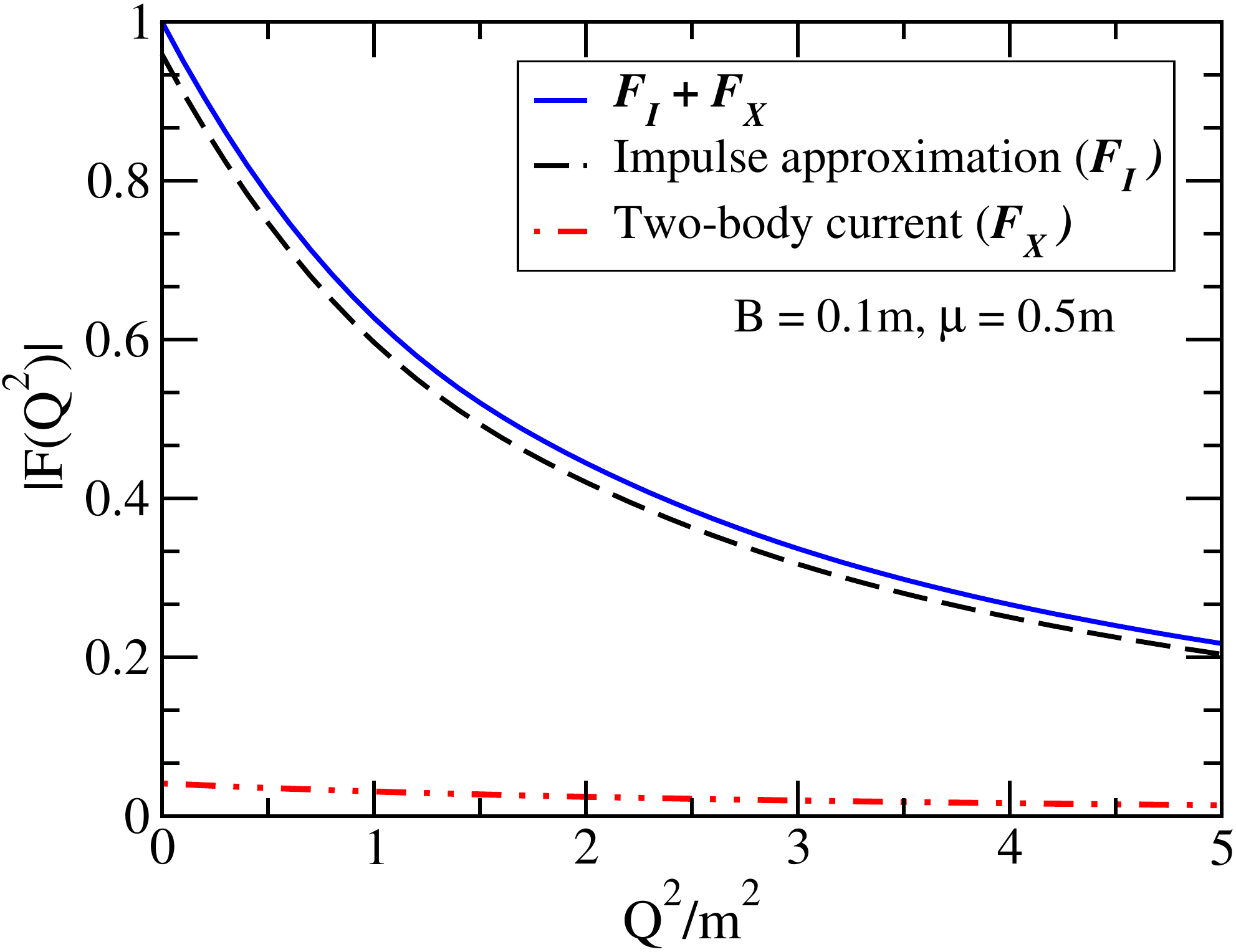}
\includegraphics[scale=0.33]{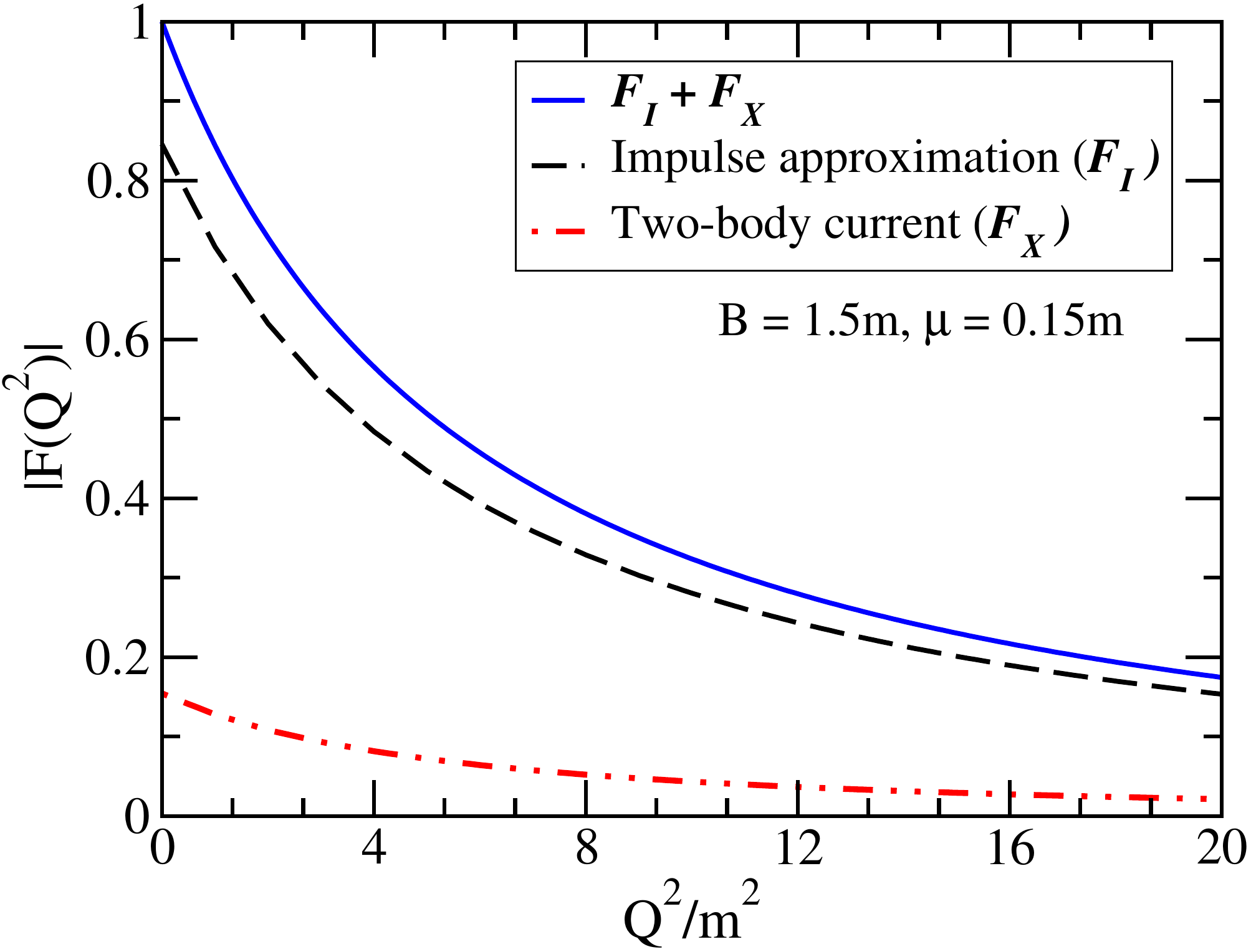}
\includegraphics[scale=0.33]{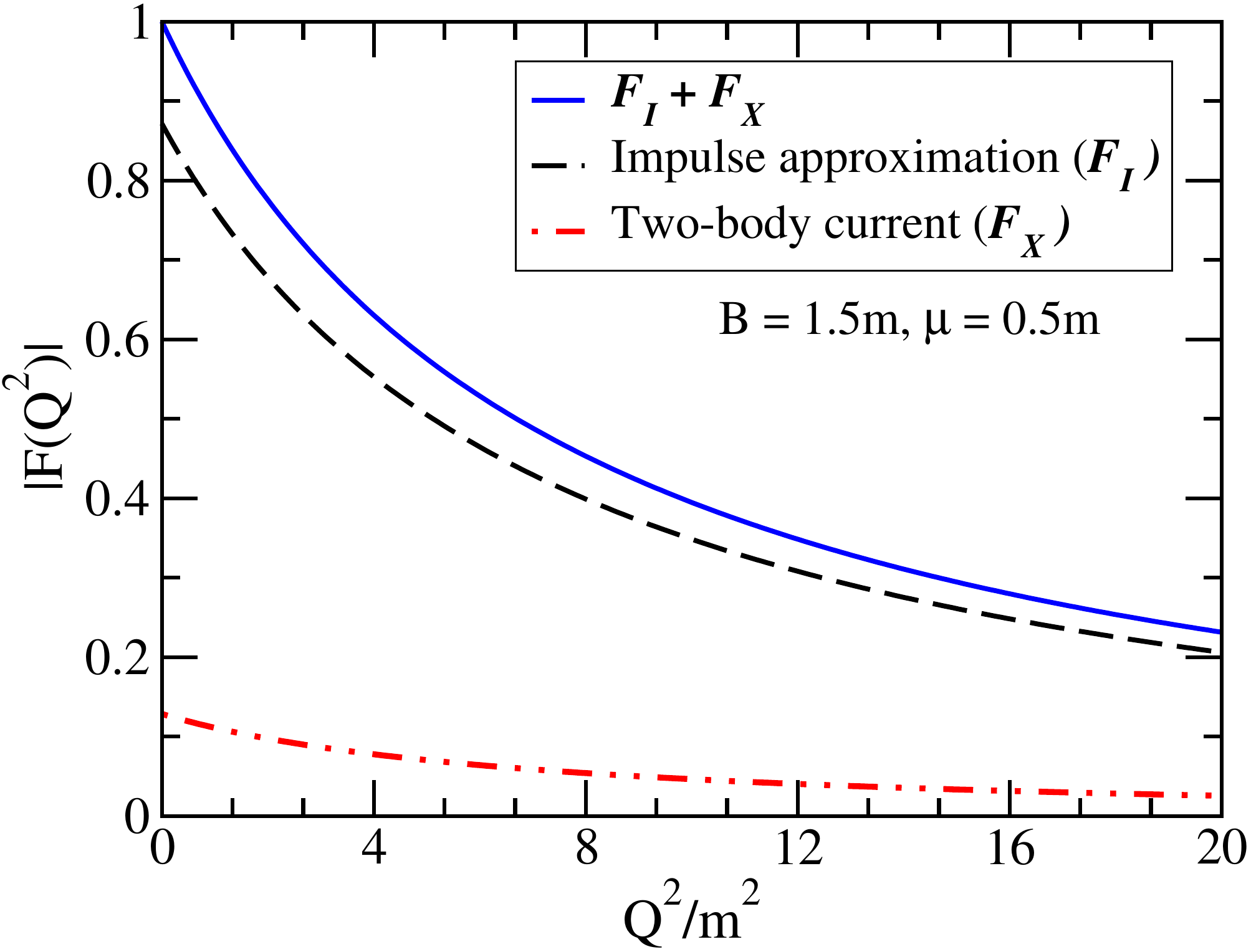}
\caption{Form factor as a function of $Q^2$. Calculations performed with the BS amplitude from the ladder plus cross-ladder kernel. The solid curve  is the full form factor. The dashed curve is the impulse contribution~($F_I$). The double-dotted dashed curve is the two-body current~($F_X$) contribution to the EM vertex. Results for:
$B=0.1\, m$ and $\mu=0.15\,m$ (upper-left frame), $B=0.1\, m$ and $\mu=0.5\,m$ (upper-right frame), $B=1.5\, m$ and $\mu=0.15\,m$ (lower-left frame), $B=1.5\, m$ and $\mu=0.5\,m$ (lower-right frame).} \label{fig:ff1}
\end{figure}

\subsection{ Results for the impulse and two-body current form factors} 

In Fig.~\ref{fig:ff1}, we present the impulse ($F_I$) and two-body current contributions ($F_X$) to the form factor, diagrammatically depicted in
Fig.~\ref{triangle}, and computed with Eqs.~(\ref{ffM}) and (\ref{formfac-5}), respectively. 
The calculations are performed for two representative 
binding energies $B= 0.1 \, m$ and $1.5 \, m$, namely weak and strong binding cases, respectively.
For both cases, the calculations are carried out with the  BS amplitude obtained with the ladder plus cross-ladder kernel 
in Eq.~(\ref{bsnew}) for exchanged boson mass of $\mu=0.15 \, m$ and $\mu=0.5 \, m$. The solid curve everywhere is the total 
form factor, normalized to one at $Q^2=0$. The total form factor is the sum of $F_I$ (dashed curve) and the $F_X$ 
(double dot-dashed curve) contributions to the 
EM vertex. We see that the relative  contribution  of $F_X$
increases when $\mu$ decreases for a given $B$, 
as the overlap between the two-body current operator and the BS amplitude increases, once
the size of the state is fixed essentially by $B$. 
The same
reason explains that by increasing the binding energy the magnitude of the contribution of the two-body current to the form
factor increases. Indeed,
in the case that we presented the maximal contribution is at $Q^2=0$ of $F_X$ (about 15\% from the total form factor) achieved for $\mu=0.15 \, m$
and $B=1.5 \, m$. This indicates that the two-body current operator contributes to short distance physics, as one could expect.

Another
feature one can extract by inspecting Fig.~\ref{fig:ff1}, is the role of the ladder exchange in shaping the large momentum region of $F_I$ 
for $Q^2>\mu^2\, , m^2$ (later on we will discuss in more details the asymptotics of the form factors). For a given binding energy, 
the change of $\mu$ modifies considerably the form factor, which  essentially is dominated by the impulse contribution
for the large momentum region, as for instance, in the case of $B=0.1 \, m$, presented in the upper frames
of Fig.~\ref{fig:ff1}. In addition, the dominance of the ladder exchange in forming the tail of the form factor is evident, and
for $Q^2/m^2=20$ one sees a scaling with $\alpha$, which changes by about a factor of about two 
when $\mu$ goes from $0.15 \, m$ to $0.5 \, m$ (see Table 1). 
This feature at large momentum is independent on the binding energy, as is exemplified, 
in the lower frames of Fig.~\ref{fig:ff1} for $B=1.5 \, m$. 
The same property is found at large transverse momentum for the LF wave function as given by Eq.~(\ref{cxi}). It is important to point out that the 
binding energy is fixed, which shapes the low momentum region of the wave function, and also to some extent the form factor.

\begin{figure}[!hbt]
\centering
\includegraphics[scale=0.33]{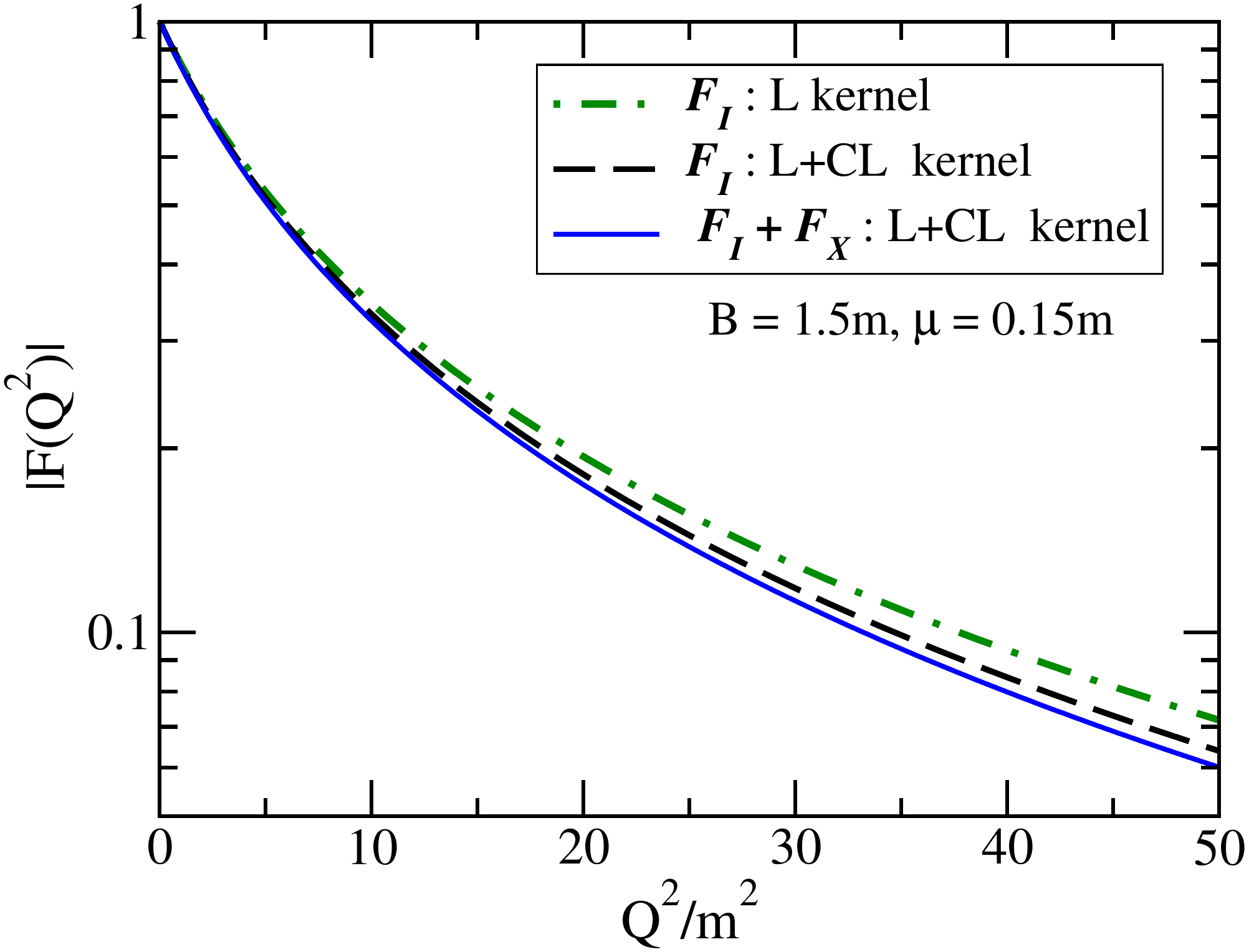}
\includegraphics[scale=0.33]{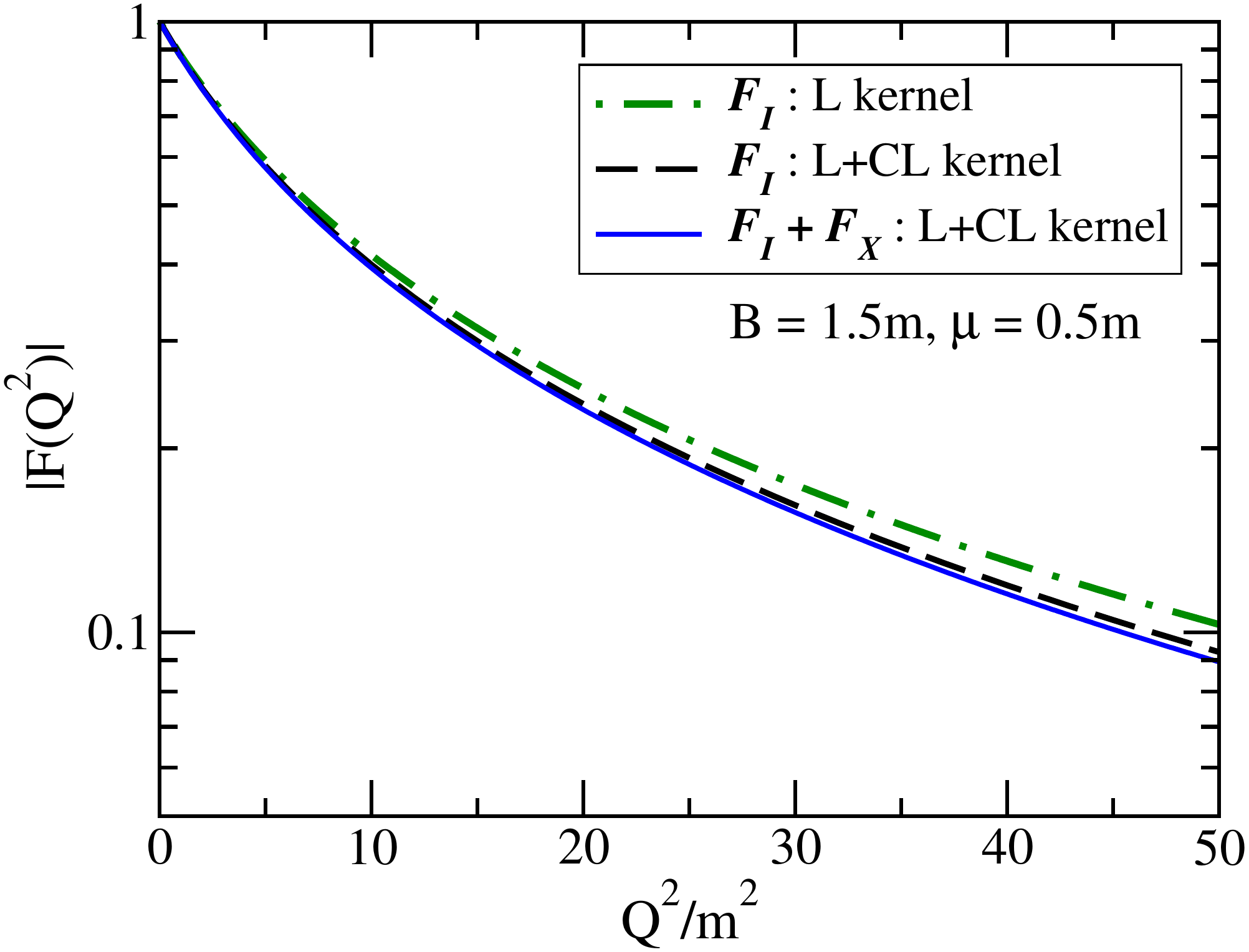}
\caption{Form factor as a function of $Q^2$. The dot-dashed curve is the form factor calculated
 with the BS amplitude found for ladder (L) kernel. The dashed curve is the impulse contribution to the form factor computed  with
the BS amplitude obtained with the ladder plus cross-ladder (L + CL) kernel. 
The solid curve is the full form factor obtained from the BS amplitude calculated with
L + CL kernel. The binding energy is $B=1.5\,m$, with the mass of the exchanged boson $\mu=0.15\,m$ (left-frame) 
and $\mu=0.5\,m$ (right-frame). All curves are normalized to 1 at $Q^2=0$.}\label{fig:ff2}
\end{figure}

In Fig.~\ref{fig:ff2} we study the sensitivity of the form factor to the dynamics, namely 
using the BS amplitude computed with ladder or with ladder plus cross-ladder kernels, for fixed binding energy 
of $B=1.5 \, m$ and exchange boson masses of $\mu=0.15 \, m$ and $ 0.5 \, m$. We choose the
 momentum transfer interval of $0\leq Q^2/m^2 \leq 50$.
We start by comparing the ladder results for the form factor with the one obtained for the
ladder plus cross-ladder kernel, considering the full current in both cases. We observe at low momentum, below
$m$, very similar slopes, that reflects close charge radius and the bound state size,
which is determined by the same binding energy. This finding is independent of 
the mass of 
the exchanged boson, as one can verify by inspecting the right and left
panels of Fig.~\ref{fig:ff2}. Comparing both frames, one observes that while the slopes are similar
by changing $\mu$, the form factor at large momentum approximately scales with $\alpha$, which we have already discussed together 
with Fig.~\ref{fig:ff1}, and the dominance of the ladder exchange in the structure of the state at large momentum.

We also compare the impulse contribution for the ladder and ladder plus cross-ladder kernels in Fig.~\ref{fig:ff2}, 
which are represented by the dot-dashed lines. For  that purpose both are normalized to one 
at zero momentum transfer. We notice two interesting features: ({\it i}) the slope is the same for
$Q\lesssim m$; ({\it ii}) at large momentum the inclusion once properly normalized the impulse contribution dominates. 
The first point ({\it i}), comes from the fact that the binding energy essentially fixes the structure at low momentum, 
the second point ({\it ii}), comes from the fact that the two-body current decreases 
much faster than the impulse contribution, as the former is a higher-twist contribution to the photon absorption  process. Indeed for large 
momentum the two-body current decays as $Q^{-2}$ with respect to the impulse contribution, which will be shown in detail in what follows.

\section{Asymptotic behavior of the form factor} \label{sect:ASYFF}

The leading behavior of the impulse and two-body current contributions to the 
form factors for $Q^2\to\infty$ can be  obtained by using standard counting rules \cite{LepPRD80}. In order
to find the leading power law behavior of the form factors represented in Figs.~\ref{fig:ff1} and \ref{fig:ff2}, 
one has to count the number of propagators, 
in which the large virtual photon momentum flows between the emission and absorption by the constituents in the bound state. 
This counting is provided, of course, by our formalism and it   results in
\begin{equation}\label{ffas}
F_I(Q^2)\sim Q^{-4} \,\,\,\text{  and  }\,\,\, F_X(Q^2)\sim Q^{-6},   
\end{equation}
apart from logarithmic corrections (see e.g. \cite{HwaNPB04}).
The two-body current is identified with a higher twist contribution and decreases faster than the impulse term by 
a $Q^{-2}$ factor. To illustrate in a transparent and analytical way how such asymptotic behavior of 
the form factors arises, we analyze it using directly Eqs.~(\ref{J}) and (\ref{formfac-2}) in the following.

\subsection{$F_I(Q^ 2)$ at large $Q^2$} 

We work in the Breit reference frame, where $\vec{p}=-\vec{p'}\equiv\vec{n}p_v$, $p_0=p'_0=\sqrt{M^2+p_v^2}$,\, $Q^2=-(p'-p)^2=
4p_v^2$  and $\vec{n}$ is the direction of the incident momentum $\vec{p}$.  Hence $p_v=\frac{1}{2}Q$, $p_0=p'_0=\sqrt{M^2+\frac{1}{4}Q^2}$
in Eq. (\ref{J}). We also denote $|\vec{k}|=k_v$. Substituting these expressions into the functions ${D}(\gamma,z;\frac{p}{2}-k,p)$ appearing in the denominator of Eq.~(\ref{J}), at large $Q$ we get:
\begin{equation}
{D}(\gamma,z;\frac{p}{2}-k,p)\approx 
(k_0-\vec{n}\cdot\vec{k})(1+z)\frac{Q}{2}+\gamma-k_0^2+k_v^2+m^2-i\epsilon -\frac{1}{2}M^2(1+z),
\end{equation}
and similarly for ${D}(\gamma',z';\frac{p'}{2}-k,p')$. Omitting a factor, we can represent the denominators in (\ref{J}) as:
\begin{equation}\label{DDp}
{D}(\gamma,z;\frac{p}{2}-k,p)\propto (1+z)Q+\delta,\quad  {D}(\gamma',z';\frac{p'}{2}-k,p') \propto (1+z')Q+\delta',
\end{equation}
where $\delta,\,\delta'$ do not depend on $Q$. Hence
\begin{multline}\label{J2}
F_{I}(Q^2)\propto
\int_{-1}^1\frac{g(z)dz}{[(1+z)Q+\delta]^3}\int_{-1}^1\frac{g(z')dz'}{[(1+z')Q+\delta']^3}=\\ =
\frac{1}{Q^6}\int_{-1}^1\frac{g(z)dz}{\left(1+z+\frac{\delta}{Q}\right)^3}\,\int_{-1}^1\frac{g(z')dz'}{\left(1+z'+\frac{\delta'}{Q}\right)^3}\,,
\end{multline}
where the variables $\gamma,\gamma'$ and the integration over $k$ are omitted since they give a finite corrections making no influence on the 
asymptotic behavior of the form factor.

If we put $\frac{\delta}{Q}=0$ in Eq.~(\ref{J2}), we get a divergent integral at $z=-1$. This means that the 
decreasing of the factor $\frac{1}{Q^6}$ can be compensated by an increasing of the values of the integrals 
at finite $Q^2$. Indeed, for $g(z)\equiv 1$, the integral  has the form:
\begin{equation}
\int_{-1}^1\frac{dz}{\left(1+z+\frac{\delta}{Q}\right)^3}\sim \frac{Q^2}{2\delta^2}.
\end{equation}
For $g(z)\equiv 1$ it gives the asymptotic form factor as $F_I(Q^2)\propto 1/Q^2$. However, the function $g(z)$ tends linearly 
to zero as $z\to -1$: $g(z)\sim (1+z)$. This weakens the compensation, therefore:
\begin{equation}
\int_{-1}^1\frac{g(z)dz}{\left(1+z+\frac{\delta}{Q}\right)^3}=\int_{-1}^1\frac{(1+z)dz}{\left(1+z+\frac{\delta}{Q}\right)^3}
\sim \frac{2Q}{\delta}.
\end{equation}
This provides the asymptotic behavior:
\begin{equation}\label{fia}
F_{I}(Q^2)\propto Q^{-4}.
\end{equation}
We can summarize the origin of this result as follows: the denominator of each propagator, containing $p$ or $p'$, according to (\ref{DDp}), contributes the factor $\sim Q$ (if the limit $Q\to\infty$ does not create a divergence).  In (\ref{J}) we have two such propagators, each in 3rd degree ($\sim \frac{1}{D^3{D'}^3}$). This gives the factor $\sim\frac{\delta^6}{Q^6}$ in (\ref{J2}). In the case of divergence (at $Q\to \infty$), like 
\begin{equation}
\int_{-1}\frac{dz}{(1+z)^2}=\left.-\frac{1}{1+z}\right|_{z\to -1},
\end{equation}
large but finite value of $Q$ eliminates the divergence, automatically replacing the limit $z=-1$ by the cutoff $z=-1+\frac{\delta}{Q}$. 
The integral becomes to be finite but large: $\sim \frac{Q}{\delta}$. In  (\ref{J2}) the product of two such integrals results in the factor 
$\sim \frac{Q^2}{\delta^2}$. This weakens the falloff  $\sim \frac{1}{Q^6}$ up to $F(Q^2)\propto \frac{1}{Q^4}$ in (\ref{J}).

Except for the term $\sim \log\left(\frac{Q^2}{m^2}\right)$ (which is out of the precision of this consideration) the asymptotic behavior 
$F_I(Q^2) \propto \frac{1}{Q^4}$ coincides with the form factor fall-off found in Eq.~(28) of Ref. \cite{HwaNPB04} for the Wick-Cutkosky model. 
The agreement of Eq.~(\ref{fia}) with the asymptotic behavior found in \cite{HwaNPB04} confirms the validity of the present consideration. 
Below we will apply this method to the two-body current contribution.

\subsection{ $F_X(Q^2)$ at large $Q^2$} 

The two-body current to the EM form factor is shown in Fig.~\ref{triangle}.
As independent integration variables we chose the four-momenta $p_2,p_8,p_9$. The other momentas are expressed as:
$p_1=p-p_2$, $p_5=p'-p_9$, $p_6=p_2-p_8$, $p_7=p_9-p_8$, $p_3=p-p_2-p_9+p_8$, $p_4=p'-p_2-p_9+p_8$.
The arguments of the BS amplitudes are: $k=\frac{1}{2}(p_1-p_2)=\frac{1}{2}p-p_2$,  $k'=\frac{1}{2}(p_5-p_9)=\frac{1}{2}p'-p_9$.
Then the two-body current in the form factor is given by Eq.~(\ref{formfac-2}).
It should be noticed that the arguments of the BS amplitudes in this equation are  $\frac{p}{2}-p_2$ and  $ \frac{p'}{2}-p_9$.  
At large $Q^2$ we omit the factors $(p'+p)^2$ and $[ (p'+p)^2 - 2(p+p')\cd ( p_9 + p_2 - p_8)] $. We can also omit the propagators
carrying the momenta $p_6$, $p_7$ and $p_8$.

The first two (cubic) factors in (\ref{formfac-2}) coming from the NIR of the two BS amplitudes 
have the same form as the corresponding factors in 
(\ref{J}). Applying to them the analysis performed for $F_I$, we find that the product of them results in $\sim \frac{1}{Q^4}$.
 However, Eq.~(\ref{formfac-2}) contains two additional propagators with $p$ and $p'$, associated with $p_3$ and $p_4$. 
 They result in an 
asymptotic behavior similar to (\ref{DDp}), but without the factors $(1+z)$, $(1+z')$. Hence, each of them adds one extra factor $\frac{1}{Q}$. They 
together give two extra powers of momentum $\sim \frac{1}{Q^2}$. Hence, the degree $\frac{1}{Q^4}$ is replaced by $\sim \frac{1}{Q^6}$. 
We conclude that the two-body current has the asymptotic behavior $F_{X}(Q^2)\propto Q^{-6}$, consistent with the counting rules. We stress that the 
asymptotic forms depend crucially on the end-point behavior of the
weight function, which is immediately  translated to the valence wave function, as seen in Eq.~(\ref{LFWF}). 

\begin{figure}[!hbt]
\centering
\includegraphics[scale=0.35,angle=0]{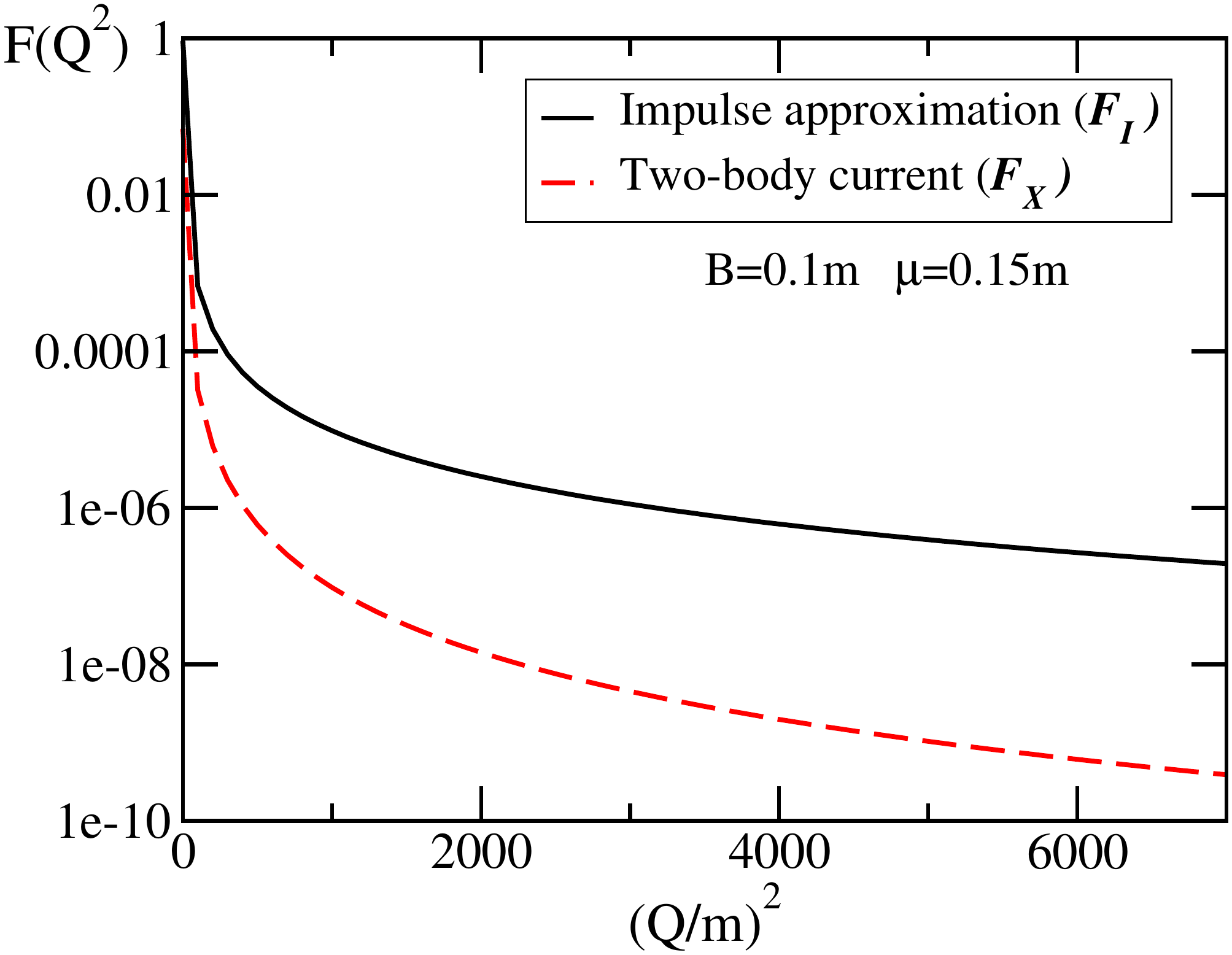}
\includegraphics[scale=0.35,angle=0]{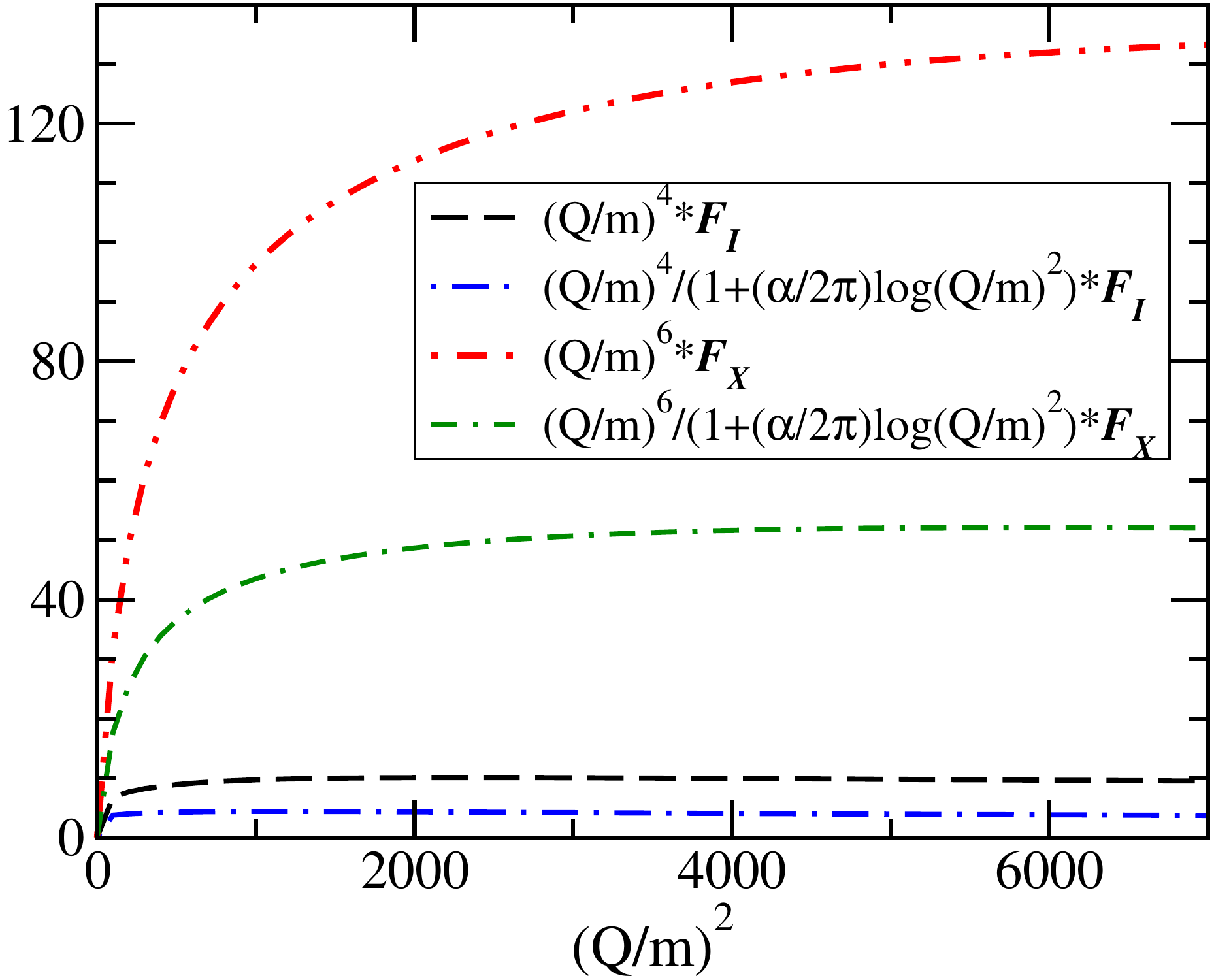}
\caption{EM form factor for the case $\mu=0.15\,m$ and  $B=0.1\,m$ obtained with the ladder plus cross-ladder kernel.
 In the left-frame the two 
contributions of the form factor are displayed. In the right-frame the asymptotic behaviors of the corresponding contributions are analyzed.} \label{fig:ff3}
\end{figure}

\subsection{Form factors at large $Q$: some numerical results}

The asymptotic behavior of the form factors given in (\ref{ffas}) are illustrated in Fig.~\ref{fig:ff3}. 
The calculations are done with the ladder plus cross-ladder kernels with $\mu=0.15\,m$ and $B=0.1\,m$. 
The results for $F_I$ and $F_X$ normalized according to Fig.~\ref{fig:ff1} are shown. 
We perform an extensive exploration for very large  momentum transfers  to check as well the leading log corrections to the form factors.
We have not derived these corrections for the form factor and just use it as suggested from the Wick-Cutkosky model, as derived in \cite{HwaNPB04}. 
We perform four studies devoted to single out the asymptotic behavior of $F_I$ and $F_X$:
({\it i}) $Q^4 \, F_I$, ({\it ii}) $Q^6 \, F_X$, ({\it iii}) $Q^4 /\left[1+(\alpha/2\pi)\log(Q/m)^ 2\right]\, F_I$, and({\it iv})
$Q^6/\left[1+(\alpha/2\pi)\log(Q/m)^ 2\right]\, F_X$.
We first observe that the asymptotic region is established for $Q/m\sim 30$, which seems reasonable as all involved scales, 
masses and binding energy are of order $m$. Second, the  products ({\it i}) and ({\it ii}) are slowly decreasing, 
while (iii) and (iv), with the inclusion of the leading log  correction, which we can distinguish in so large momentum transfer interval presented in Fig.~\ref{fig:ff3}, show an improvement in getting the flat behavior 
at large momentum.

\section{ Summary and outlook} \label{sect:SUMOUT}

The response of the Minkowski space structure of a two-boson bound state,
within a Yukawa model with a scalar boson exchange, to the inclusion of 
the cross-ladder contribution to the ladder kernel of the BS equation was investigated quantitatively. 
The NIR allied with the LF projection was used to solve numerically the 
BS equation in Minkowski space.
We computed both the valence wave function and elastic electromagnetic form factor 
 including the two-body current contribution to the electromagnetic vertex.
We have discussed in 
detail the dependence on the ladder exchange in  building the asymptotic behavior of the valence wave function and form factor, 
for a fixed binding energy, considering both ladder and ladder plus cross-ladder kernels. This allowed us to single out the dominance of the
ladder exchange, by comparing results for a fixed binding energy and using the two interacting  kernels. 

The valence wave function at low transverse momentum is independent of the kernel, being 
determined just by the given binding energy. We also studied quantitatively 
the factorization of the valence wave function in terms of the transverse  and longitudinal momenta at large transverse
momentum \cite{GutPLB16}. In this case, as expressed by Eq.~(\ref{cxi}), once $\alpha$ is factorized out
 for both binding energy and normalization fixed, the form of the wave function with the 
 longitudinal momentum fraction is quite universal. In the case of
$B/m > 0.1$, we found that the functional form approaches  the Wick-Cutkosky solution $[\xi(1-\xi)]^2$ obtained for $B=2m$ 
in Ref.~\cite{HwaNPB04}. Our conjecture is that the form and magnitude of $C(\xi)$ and the wave function at low transverse momentum, for the 
normalization $\psi_{LF}(0,1/2)=1$ and a given binding energy, are to great deal independent on the inclusion of the irreducible cross-ladder 
contributions in higher order in the kernel.
Then, we can turn our attention to the nice work \cite{NiePRL96}, where the problem with the generalized ladder kernel
 was solved by means of the Feymman-Schwinger representation. Making use of that, one can speculate on the 
 form of the valence wave function when an infinite set of cross-ladder diagrams are included in the kernel.

The electromagnetic 
current in the case of the cross-ladder kernel includes, besides the impulse term, a two-body current obtained by gauging the cross-ladder
kernel. We note that due to the symmetry of the elastic virtual photo-absorption amplitude, the impulse and 
two-body amplitudes, conserve current independently, which is not the case in an inelastic transition.
Our numerical results show that for a given binding energy, the two-body current becomes more relevant as 
lighter is the exchanged boson mass as well as when the binding energy becomes larger. This is easy to understand 
if one considers that in both cases the overlap between the bound state and the two-body current increases, either by increasing the range 
of the interaction or decreasing the size of the bound state. For zero momentum transfers, where the two-body 
current is more relevant, and for a strongly bound system the contribution is about 15 $\%$ of the normalization. 

The form factor in the large momentum region was studied in detail and the power-law decreasing, as expected from the counting 
rules
applied to our model, was derived using the adopted Nakanishi integral representation of Bethe-Salpeter amplitude. The leading
contribution to the dependence on the large momentum transfer comes from the ladder exchange, which was
illustrated by comparing the impulse term from ladder and ladder plus cross-ladder kernels, 
where the proportionality of the tail to $\alpha$ was singled out for a fixed binding energy. 
It was pointed out the crucial role of the end-point behavior of the Nakanishi weight function in the power-law behavior.

Although, the present study is focused on the two-boson problem, the present analysis can be extended to fermionic systems,
for which the Bethe-Salpeter amplitude has been obtained by means of the Nakanishi representation \cite{CarEPJA10,dPaPRD16}. 
It has of course  wide applications to the study of meson structure. 

{\it Acknowledgments.} We thank the support from Conselho Nacional de Desenvolvimento Cient\'ifico e Tecnol\'ogico(CNPq) and Coordena\c c\~ao de Aperfei\c coamento de Pessoal de N\'ivel Superior 
(CAPES) of Brazil. J.H.A.N. acknowledges the support of the grant \#2014/19094-8 and V.A.K. of the grant \#2015/22701-6 from Funda\c c\~ao
de Amparo \`a Pesquisa do Estado de S\~ao Paulo (FAPESP).

\end{document}